\documentstyle[preprint,aps]{revtex}
\begin{document}

\draft   


\title{Localization of Growth Sites in DLA Clusters:\\
Multifractality and Multiscaling}

\author{Jysoo~Lee$^{1}$,~Stefan~Schwarzer$^2$,~Antonio~Coniglio$^{2,3}$~and~H.~Eugene~Stanley$^2$}

\address{$^1$ HLRZ-KFA J\"ulich, Postfach 1913, W-5170 J\"ulich, Germany}
\address{$^2$ Center for Polymer Studies and Department of Physics\\
Boston University, Boston, MA 02215, USA}
\address{$^3$ Dipartimento di Scienze Fisiche\\
Universita degli Studi di Napoli, I-80125 Napoli, Italy}

\date{\today}

\maketitle

\begin{abstract}
The growth of a diffusion limited aggregation (DLA) cluster with mass $M$
and radius of gyration $R$ is described by a set of growth probabilities
$\{ p_i\}$, where $p_i$ is the probability that the perimeter site $i$
will be the next to grow.  We introduce the joint distribution $N(\alpha,
x, M)$, where $N(\alpha,x,M)d\alpha dx$ is the number of perimeter sites
with $\alpha$-values in the range $\alpha\le \alpha_i \le \alpha+d\alpha$
(``$\alpha$-sites'') and located in the annulus [x, x+dx] around the
cluster seed. Here, $\alpha_i \equiv -\ln p_i / \ln R$ if $p_i>0$,
$x\equiv r_i/R$, and $r_i$ is the distance of site $i$ from the seed of
the DLA cluster.  We use $N(\alpha,x,M)$ to relate multifractal and
multiscaling properties of DLA.  In particular, we find that for large
$M$ the location of the $\alpha$-sites is peaked around a fixed value
$\bar x(\alpha)$; in contrast, the perimeter sites with $p_i=0$ are
uniformly distributed over the DLA cluster.
\end{abstract}


\narrowtext

\section{Introduction}
\label{s:introduction}

The growth of a diffusion limited aggregation (DLA)
\cite{Witten81,Turkevich85,Ball85,Coniglio85,Parisi85,Hayakawa87,Halsey87,Pietronero88,Meakin88,b:Feder88,b:Stanley88,b:Pietronero90,b:Bunde91,b:Vicsek92}
cluster with mass $M$ is described by the set of growth probabilities
$\{p_i\}$ \cite{Meakin85,Meakin86,Halsey86,Amitrano86} where $p_i$ is the
probability that perimeter site $i$ will be the next to grow. One way to
analyze the set $\{p_i\}$ is by calculating the ``growth-probability
distribution'' $n(\alpha,M)$, where $n(\alpha, M) d\alpha$ is the number
of perimeter sites with $\alpha \le
\alpha_i \le \alpha+d\alpha$,
\begin{equation} \label{e:alpha}
   \alpha_i \equiv -\ln p_i / \ln R,
\end{equation}
and $R$ is the radius of gyration of the cluster
\cite{Amitrano86,Amitrano89}.  We call $\alpha$-sites those sites which
are characterized by the same value of $\alpha$. The main motivation for
studying the distribution $n(\alpha,M)$ is its relation to the
multifractal spectrum $f(\alpha)$
\cite{Halsey86a,Stanley88} of the ``measure''
$\{p_i\}$. We define $f(\alpha,M)$ through
\begin{equation} \label{e:falpha}
   n(\alpha,M) \equiv M^{\nu f(\alpha,M)},
\end{equation}
where $\nu$ is the inverse fractal dimension of DLA, $R\sim M^\nu$.  If
for large $M$ $f(\alpha,M)$ converges to an $M$-independent function
$f(\alpha)$, then $f(\alpha)$ is usually called the multifractal
spectrum. For 2D DLA, there exist several studies
\cite{Mandelbrot90,Mandelbrot91,Schwarzer91}
proposing different convergence behaviors and functional forms of
$f(\alpha)$ in the limit $M\to\infty$. However, these considerations will
not be essential for our arguments concerning the relation of
multiscaling and multifractality. Henceforth we will only assume that
some $f(\alpha)$ exists.

During the process of calculating $n(\alpha,M)$, the information
about the location of the $\alpha$-sites is lost. However, some
information about the location of the growth sites and their associated
values $p_i$ may be obtained from the Plischke-R\'acz probability
$P(x,M)$ \cite{Plischke84}, where $P(x,M)dx$ is the probability that the
next particle will be deposited at a rescaled distance $x \le x_i \le
x+dx$.  Here $x_i\equiv r_i/R$ and $r_i$ is the distance from the
cluster seed.  For DLA, the function $P(x,M)$ displays a peak at a
constant value $\bar x$ of the deposition radius \cite{Plischke84}.

A simple form for $P(x,M)$ is the Gaussian,
\begin{equation} \label{e:pxm-gaussian}
  P(x,M) = {1\over\sqrt{2\pi\xi^2}}
    \exp\left[ -{ {(x-\bar x)^2}\over 2\xi^2}\right],
\end{equation}
where $\xi^2$ denotes the mean square width of the deposition zone.
Plischke and R\'acz \cite{Plischke84} suggest that $\xi \sim M^{\nu'-\nu}$,
$\nu'<\nu$, where $\nu$ is the inverse fractal dimension of DLA, $R\sim
M^\nu$, and $\nu'$ an independent exponent.  However, Meakin and Sander
\cite{Meakin85a} find that $\nu'$ approaches $\nu$ as $M$ increases. They
also argue that $\nu = \nu'$ in the limit $M\to\infty$.

If $(\bar x /\xi)^2 \simeq 2c\ln M$ \cite{Coniglio90} with constant
$c$, then $P(x,M)$ takes the form
\begin{equation} \label{e:pxm-multiscaling}
  P(x,M) = {1\over\sqrt{2\pi\xi^2}} M^{-\phi(x)} C_{\rm PR}(x),
\end{equation}
where from (\ref{e:pxm-gaussian}) $\phi(x) = c (x/\bar x -1)^2$ and
$C_{\rm PR}(x) = 1$. In general, if $C_{\rm PR}(x)$ is a generic function of
$x$ and
$\phi(x)$ is independent of $x$ we have conventional scaling while if
$\phi(x)$ is $x$-dependent we say that we have {\it multiscaling.}


Next, we introduce the annular density $\rho_A(x,M)$, where
$\rho_A(x,M)dx$ is the number of particles in an annulus $[x,x+dx]$ and
related to the conventional particle density $\rho(r,M)$ by
$\rho_A(x,M)dx = 2\pi r\rho(r,M)dr$.  Since the change of $\rho_A(x,M)$
with increasing cluster mass is given by $P(x,M)$
\cite{Plischke84}, i.e.,
\begin{equation} \label{e:rho-derivative}
{\partial \over \partial M}\rho_A(x,M) = P(x,M),
\end{equation}
it was suggested \cite{Amitrano91} to write $\rho_A(x,M)$ in the same
multiscaling form as Eq. (\ref{e:pxm-multiscaling}),
\begin{equation} \label{e:rho-multiscaling}
 \rho_A(x,M) = r^{D(x)}C_\rho(x),
\end{equation}
where $D(x)$ is the fractal dimension for a thin annulus with average
radius $x$, and $C_\rho(x)$ is an amplitude.
%
%
Note that if $\phi=\phi(x)$ in Eq.
(\ref{e:pxm-multiscaling}), then also $D=D(x)$ in
(\ref{e:rho-multiscaling}).  Multiscaling in (\ref{e:rho-multiscaling})
has been supported by simulations \cite{Amitrano91}. Whereas
multiscaling for clusters with $M < 10^6$ has been confirmed very
recently by P. Ossadnik \cite{PO-private}, the same study analyzes one
very large off-lattice cluster of $M = 5 \times 10^7$ arriving at
an ambiguous result, which is consistent with both multi- and standard
scaling of $\rho_A(x,M)$.

As demonstrated in Ref. \cite{Coniglio90}, multiscaling results if the
$\alpha$-sites are ``localized' in space. Here, we study the nature of
the ``localization'' of the $\alpha$-sites and the non-localized behavior
of the $p_i = 0$ (``dead'') perimeter sites (Secs.
\ref{s:joint-distribution-function}, \ref{s:simulation}). The
consequences for the multiscaling hypothesis of $P(x,M)$ and
$\rho_A(x,M)$ will be discussed in Sec. \ref{s:cases}.  Moreover, we
introduce the notion of a multifractal spectrum in an annulus and find an
intriguing combination of multifractal and multiscaling properties (Sec.
\ref{s:multifractal}) \cite{Jensen91}.

\section{The joint distribution function}
\label{s:joint-distribution-function}

In this section we introduce the joint distribution function
$N(\alpha,x,M)$, where $N(\alpha,x,M) d\alpha dx$ is the number of
perimeter sites with $p_i > 0$ such that $\alpha\le \alpha_i \le
\alpha+d\alpha$ and within the annulus [x, x+dx].
The distribution $N(\alpha,x,M)$ can be related to the three functions
discussed in the introduction:

(a) $n(\alpha,M)$. By integration of $N(\alpha,x,M)$ over $x$, we have,
\begin{equation} \label{e:nalphaM}
   n(\alpha,M) = \int dx N(\alpha,x,M).
\end{equation}

(b) $P(x,M).$ If we use Eq. (\ref{e:alpha}) together with the relation
$R=M^{\nu}$ --- which reflects the well-established fractal structure of
DLA --- to write the growth probability as $M^{-\nu\alpha}$, then
\begin{equation} \label{e:pxm}
  P(x,M) = \int d\alpha N(\alpha,x,M) M^{-\nu\alpha}.
\end{equation}

(c) $\rho_A(x,M)$. One possibility to express $\rho_A(x,M)$ in terms of
$N(\alpha,x,M)$ is by using Eqs. (\ref{e:rho-derivative}) and
(\ref{e:pxm}). Integration of Eq. (\ref{e:rho-derivative}) with respect
to $M$ yields
\begin{mathletters}
\begin{equation}  \label{e:rho-nalpha-integral}
  \rho_A(x,M) = \int_1^M dM' \int d\alpha N(\alpha,x,M') (M')^{-\nu\alpha}
\end{equation}
However, we would like to point out another relationship that does not
involve an integration over the growth history of the cluster.  First,
note that $N(\alpha,x,M)$ only describes perimeter sites with growth
probability $p_i > 0$ (alive perimeter sites --- in contrast to dead
sites with $p_i=0$, see Fig. \ref{f:configurations}).  We denote the
annular density profile of alive perimeter sites by $\rho_A^{\rm
(a)}(x,M)$ and that of dead perimeter sites by $\rho_A^{\rm(d)} (x,M)$,
both defined in analogy to $\rho_A(x,M)$, which describes the density of
cluster sites.  The sum $\rho^{\rm (a)}_A(x,M) + \rho^{\rm (d)}_A (x,M)$
describes {\it all\/} perimeter sites of clusters of mass $M$. Since DLA
is a treelike fractal object \cite{Meakin84}, we expect the annular
density of all perimeter sites to be proportional to the annular density
of cluster sites $\rho_A(x,M)$.  Furthermore, as one can see from Figs.
\ref{f:dead-alive} and
\ref{f:distributions}, the
spatial distributions of dead and alive sites are in good approximation
proportional to each other, i.e., $\rho_A^{\rm (a)}(x,M) \sim
\rho_A^{\rm (d)}(x,M)$; a detailed discussion is given in Appendix
\ref{a:1}. Consequently, we expect
\begin{equation} \label{e:rho-nalpha}
  \rho_A(x,M) \sim \int d\alpha N(\alpha,x,M).
\end{equation}
\end{mathletters}

\section{Simulation Approach}
\label{s:simulation}

A calculation of the joint distribution $N(\alpha,x,M)$ shows that the
$x$ dependence is approximately Gaussian, centered around a value $\bar x
\equiv \langle x \rangle$ with variance $\xi^2(\alpha,M) \equiv \langle
x^2\rangle - \langle x \rangle^2$, i.e.,
\begin{equation} \label{e:naxm-gauss}
N(\alpha,x,M) \propto { 1\over \sqrt{2\pi \xi^2(\alpha,M)} }
   \exp\left[- {{(x-\bar x(\alpha))^2} \over 2\xi^2(\alpha,M)}\right].
\end{equation}
Here, the brackets $\langle \ \rangle$ indicate an expectation value with
respect to the empirical distribution $N(\alpha,x,M)$, i.e.,
$\langle f(x)\rangle \equiv \int dx f(x) N(\alpha, x, M) /
\int dx N(\alpha,x,M)$.

In Fig. \ref{f:naxm}, we show $N(\alpha, x, M)$ with $M = 20\ 000$ for
$\alpha$ values of $1.1, 1.5, 1.9, 2.3$ and $2.8$ vs. $\pm (x-\bar x)^2$,
where the $+$ sign applies if $x \ge \bar x$ and $-$ otherwise.  The
Gaussian approximation (\ref{e:naxm-gauss}) is justified, since we
observe for positive and negative abscissa values an approximate straight
line behavior of $N(\alpha,M)$.  For each $\alpha$, the modulus of the
slope $m(\alpha)$ of these lines relates to the width of the Gaussian,
$|m(\alpha)| = 1/2\xi^2(\alpha,M)$. Apparently, $\xi(\alpha,M)$ increases
with $\alpha$. In other words, highly screened growth sites are less
localized than the exposed growth sites in the active region of the
cluster that are characterized by small values of $\alpha$. The
approximation is worse for $x < \bar x$, where especially for large
values of $\alpha$ the presence of the cluster center distorts the pure
Gaussian behavior \cite{Ossadnik92a}.

In Fig. \ref{f:alpha-sites} we demonstrate that $\alpha$-sites are
located in approximate annuli around the center of the cluster by
displaying all the ``live'' ($p_i > 0$) perimeter sites of $18$
superposed off-lattice DLA clusters of $M=20\ 000$ for three distinct
values of $\alpha$. In contrast, the dead sites of a cluster, as shown
in Fig. \ref{f:dead-alive}a, are distributed with a density proportional
to that of the alive sites (cf.  App. \ref{a:1}), which, for comparison,
are displayed in Fig. \ref{f:dead-alive}b.

The $M$ dependence of the width $\xi(\alpha,M)$ is shown in Fig.
\ref{f:width-mean}a.  We cannot identify an unequivocal limit behavior as a
function of $M$.  For certain values of $\alpha$, $\xi(\alpha,M)$ seems
to decrease as $M\to\infty$ --- for others an increase is observed. The
best statistics are obtained for $\alpha \approx 1.36$, for which we find a
decreasing width.  Since asymptotically $\xi(\alpha,M)$ can certainly not
increase --- which would correspond to the statement that the growth zone
would become larger than the cluster itself --- we are clearly still in a
mass regime where finite size effects play an important role.  Thus, we
will have to discuss several possibilities for the $M$-dependence of $\xi
(\alpha,M)$ in the following sections.

The $M$ dependence of the mean $\bar x(\alpha, M)$ is plotted in Fig.
\ref{f:width-mean}b. Unlike in the case of the width $\xi(\alpha,M)$, here the
limit behavior for large $M$ is manifest.  For small values of $\alpha$,
$\bar x(\alpha, M)$ decreases with $M$ while for large $\alpha$, $\bar
x(\alpha,M)$ increases with $M$.  However, in both cases $\bar
x(\alpha,x,M)$ converges towards an $M$-independent limit $\bar
x(\alpha)$, as shown in Fig. \ref{f:width-mean}c. Note that $\bar
x(\alpha)$ is a monotonically decreasing function and thus invertible.
The decrease of $\bar x(\alpha)$ with $\alpha$ results from the stronger
screening (large $\alpha$) in the interior of the cluster (small $x$).

\newsavebox{\boxp}
\newsavebox{\boxr}
\savebox{\boxp}{$P(x,M)$}
\savebox{\boxr}{$\rho_A(x,M)$}

\section{Multiscaling of \usebox{\boxp} and \usebox{\boxr} }
\label{s:cases}

Next we discuss the functional form of the new distribution function
$N(\alpha,x,M)$.  As we have shown above, our calculations are
consistent with the possibility that for large $M$, $N(\alpha, x,M)$ is a
Gaussian \cite{generalization} in $x$,
\begin{equation} \label{e:naxm}
N(\alpha,x,M) = {M^{\nu f(\alpha)} \over \sqrt{2\pi \xi^2(\alpha,M)} }
  \exp\left[- {{(x-\bar x(\alpha))^2} \over 2\xi^2(\alpha,M)}\right].
\end{equation}
In (\ref{e:naxm}) we write $\bar x=x(\alpha)$, since we have found that
$\bar x$ for large $M$ depends only on $\alpha$. The term $M^{\nu
f(\alpha)}$ in the prefactor of the Gaussian ensures that $n(\alpha, M)$
has the multifractal form (\ref{e:falpha}), as one can see by
integration with respect to $x$.

We next discuss the implications for multiscaling of several
possibilities for the functional dependence of the width $\xi(\alpha,M)$
on $\alpha$ and $M$. Preserving the multifractal properties of the
distribution, we explore three mutually exclusive cases for the mass
dependence of $\xi(\alpha,M)$:

(A) Constant width, $\xi(\alpha,M) = A(\alpha).$ A constant width
corresponds to the possibility that both the average location of the
$\alpha$-sites and the width of the growth zone are not mass dependent,
implying that both length scales are proportional to the cluster radius
$R$.

(B) ``Strong localization,'' $\xi(\alpha,M) = A(\alpha) M^{-y}$, $y>0$,
corresponds to a fast decrease of the width of the
spatial distribution of the $\alpha$-sites. Asymptotically, the
$\alpha$-sites are located at the same distance
$x(\alpha)$ from the cluster seed.

(C) ``Weak localization,'' $\xi(\alpha,M) = A(\alpha) / \sqrt{\ln M},$
where the width still tends to zero, but in a logarithmic fashion with
intriguing implications for the scaling behavior of $P(x,M)$ and
$\rho_A(x,M)$.

\savebox{\boxp}{$\xi(\alpha,M) = A(\alpha)$}
\subsection{Constant Width: \usebox{\boxp}}
\label{s:case-i}

 Substituting relation (\ref{e:naxm}) into
(\ref{e:pxm}), we obtain for the Plischke-R\'acz probability
\begin{eqnarray} \label{e:pxm-case-i}
P(x,M) & = & \int d\alpha {M^{\nu f(\alpha)}M^{-\nu\alpha} \over
                       \sqrt {2 \pi A^2(\alpha)} }\nonumber \\
 & &\qquad\times\exp\left[- {(x-\bar x(\alpha))^2 \over 2 A^2(\alpha)}\right].
\end{eqnarray}
Performing a steepest descent analysis of (\ref{e:pxm-case-i}), we now
calculate the value $\alpha^*$ of $\alpha$ at which the integrand is
maximal. In the $M \to\infty$ limit, the resulting condition for
$\alpha^*$ is
\begin{equation} \label{e:information}
{d \over d\alpha} f(\alpha)|_{\alpha^*} = 1.
\end{equation}
The value of $\alpha^*$ which satisfies Eq. (\ref{e:information})
is known to be unity \cite{Makarov85}, moreover, for $\alpha = 1$ we
have $f(\alpha) = 1.$ Thus, in case (A), $P(x,M)$ is
$M$-independent and has Gaussian shape with constant width
$\xi = A(1)$, i.e.,
\begin{equation} \label{e:pxm-result-i}
  P(x,M) \sim {1\over\sqrt{2\pi A^2(1)} }
  \exp\left[-{ (x-\bar x(1))^2\over 2A^2(1)} \right].
\end{equation}
Eq. (\ref{e:pxm-result-i}) does not explicitly depend on $M$ and thus
$P(x,M)$ obeys standard scaling.

We perform a similar analysis to evaluate the expression
(\ref{e:rho-nalpha}) for the density profile of the cluster,
\begin{equation} \label{e:rho-case-i}
\rho_A(x,M) \sim \int d\alpha{ M^{\nu f(\alpha)}\over \sqrt{2\pi A^2(\alpha)}}
            \exp\left[-{(x-\bar x(\alpha))^2 \over 2A^2(\alpha) }\right].
\end{equation}
Here, the condition for the saddle-point value $\alpha^*$ in the
limit of large $M$ becomes,
\begin{equation} \label{e:maximum}
        {d\over d\alpha} f(\alpha)|_{\alpha^*} = 0
\end{equation}
The maximum $f(\alpha^*)$ has the value $f(\alpha^*)= 1/\nu$, so that
the density can be written as
\begin{equation} \label{e:rho-result-i}
 \rho_A(x,M) \sim {M\over \sqrt{2\pi A^2(\alpha^*)} }
        \exp\left[-{(x-\bar x(\alpha^*))^2 \over 2A^2(\alpha^*)}\right].
\end{equation}
Formally, Eq. (\ref{e:rho-result-i}) can be cast into the form of Eq.
(\ref{e:rho-multiscaling}),
\begin{eqnarray} \label{e:rho-result-i-msform}
        \rho_A(x,M) & \sim & R^{1/\nu} {r^{1/\nu} \over r^{1/\nu}}
           \exp\left[-{(x-\bar x(\alpha^*))^2 \over
                       2A^2(\alpha^*)} \right]  \cr
             & = & r^{1/\nu} C_\rho(x).
\end{eqnarray}
Note that in case (A) the exponent of $r$ is independent of $x$ and
thus also $\rho_A(x,M)$ obeys standard scaling.

It is instructive to note the meaning of the values $\alpha^*$ from
Eqs.  (\ref{e:information}) and (\ref{e:maximum}) in the multifractal
spectrum.  The probability to grow at a site with a specific value of
$\alpha$ is maximal when the product of growth probability and number of
sites with this probability, and thus $f(\alpha)-\alpha$, is maximal.
As can be seen by differentiating with respect to $\alpha$ this
condition is equivalent to relation (\ref{e:information}), from which
results the dominant value $\alpha^*$ that controls the $P(x,M)$
integral (\ref{e:pxm-case-i}). In contrast, the mass distribution
$\rho_A(x,M)$ (\ref{e:rho-case-i}) in the cluster is controlled by the
$\alpha$ value corresponding to the maximum of $f(\alpha)$, which is the
fractal dimension of the set comprising the ``majority'' of growth
sites.

\paragraph*{}
However, the simple possibility that the width $\xi(\alpha,M)$ of
$N(\alpha,x,M)$ is independent of $M$ appears questionable. First, a
recent calculation \cite{Ossadnik92a} indicates that the width of $P(x,M)$
decreases with $M$. Second, simulation results for off-lattice DLA
clusters with $M$ up to $5\times 10^7$ are consistent with
the multiscaling relation (\ref{e:rho-multiscaling}) for
$\rho_A(x,M)$ which precludes an $M$ independent width $\xi(\alpha,M)$.

\savebox{\boxp}{$\xi(\alpha,M) = A(\alpha) M^{-y}$}
\subsection{Strong Localization: $\xi(\alpha,M) = A(\alpha) M^{-y}$}
\label{s:case-ii}

The implications  for case (B), $\xi(\alpha,M) = A(\alpha)M^{-y}$,
are quite different from case (A). Now, the analog of Eq. (\ref{e:pxm-case-i})
for the Plischke-R\'acz probability becomes
\begin{eqnarray} \label{e:pxm-case-ii}
P(x,M) & = & \int d\alpha { M^{\nu f(\alpha)-\nu\alpha} \over
                       \sqrt {2 \pi A^2(\alpha) M^{-2y}} } \nonumber \\
  & & \quad\times\exp\left[- M^{2y}{(x-\bar x(\alpha))^2
                        \over 2 A^2(\alpha)}\right].
\end{eqnarray}
For sufficiently large $M$, the Gaussian in the integrand tends to a
delta function centered  at $x$,
\begin{equation} \label{e:alpha-case-ii}
     x =\bar x( \alpha ).
\end{equation}

Given an annulus $x$, there is only one value of
$\alpha = \alpha(x)$ given by the inverse function of $\bar x(\alpha)$.
 --- the monotonicity of $\bar x(\alpha)$ guarantees
the existence of $\alpha(x)$ (cf. Sec. \ref{s:simulation}
and Fig. \ref{f:alpha-sites}).

Using (\ref{e:alpha-case-ii}), we can compare $P(x,M)$, Eq.
(\ref{e:pxm-case-ii}), to the multiscaling form
(\ref{e:pxm-multiscaling}) and find that
\begin{equation} \label{e:pxm-result-ii}
  \phi (x) =  - \nu f(\alpha(x)) + \nu \alpha(x).
\end{equation}
Thus, in contrast to case (A), case (B) results in multiscaling.

In the same fashion, from (\ref{e:rho-nalpha-integral}) and
(\ref{e:naxm}) we can
demonstrate multiscaling for the density profile of the cluster.  Again
the gaussian in (\ref{e:naxm}) tends to a delta function centered at
$x=\bar x(\alpha)$ and thus (\ref{e:alpha-case-ii}) also determines
$\alpha(x)$ for the density profile.  The resulting density profile
$\rho_A(x,M)$ is
\begin{equation} \label{e:rho-result-ii}
 \rho_A(x,M) \sim M^{\nu f(\alpha)}
             \sim r^{f(\alpha(x))} C_\rho(x),
\end{equation}
and displays multiscaling as in Eq. (\ref{e:rho-multiscaling}) with
\begin{equation}
    D(x) = f(\alpha(x)),
\end{equation}
in agreement with the result of Ref \cite{Coniglio90}.

The results (\ref{e:pxm-result-ii}) and (\ref{e:rho-result-ii})
are not altogether surprising.
In fact, the distribution $N(\alpha,x,M)$ for fixed
$\alpha$ tends to a $\delta$-function as $M\to\infty$.
In this limit, almost all the
sites with a specific $\alpha$-value are located at distance $\bar x
(\alpha)$ from the cluster seed --- such that we refer to case (ii) as
``strong localization.'' Vice versa, a specific location $x$ singles out
an $\alpha$ value $\alpha(x)$. From $f(\alpha)$ we then obtain the
fractal dimension of the set of these $\alpha(x)$-sites.
Eq. (\ref{e:rho-result-ii}) can now be understood just as the
usual relationship between mass and extension of a fractal object and,
similarly, $M^{-\phi(x)}$, which describes the
probability of deposition at $x$, is just the product of the growth probability
$M^{-\nu\alpha(x)}$ at $x$ and the multiplicity of the growth sites at
$x$, $M^{\nu f(\alpha(x))}$.

In the case of strong localization it is particularly simple to obtain
an understanding of the relationship between cluster structure and the
distribution of growth probabilities.  The large $\alpha$ part of
$f(\alpha)$ reflects the properties of the frozen region (small $x$) of
the cluster, where the $p_i$ are so small that effectively no further
growth will occur. One expects that the mass distribution in the frozen
region is characterized by the fractal dimension of DLA. This assumption
is supported by the results of Refs.
\cite{Amitrano91,PO-private,Ossadnik92a}.
However, since $\alpha(x)$ is not constant for small $x$, and if our
assumption (\ref{e:naxm}) for the form of the joint distribution
function $N(\alpha,x,M)$ is still valid, then $f(\alpha)$ has to be {\it
independent of $\alpha$\/} for large $\alpha$. In fact, the phenomenon
of a ``phase transition'' \cite{phase-transition,Lee89} in DLA is
consistent with such a behavior of $f(\alpha)$ \cite{Mandelbrot90}.

\savebox{\boxp}{$\xi(\alpha,M) = A(\alpha) / \sqrt{\ln M}$}
\subsection{Weak Localization: \usebox{\boxp}}
\label{s:case-iii}

For case (C), $\xi(\alpha,M) = A(\alpha) \sqrt{\ln M}$.
The exponential in the integrands of (\ref{e:pxm-case-i})
and (\ref{e:pxm-case-ii}) turns into a power-law,
\begin{equation} \label{e:pxm-case-iii}
 P(x,M) =
 \int d\alpha \sqrt {\ln M \over 2\pi A^2(\alpha)} M^{\nu f(\alpha)
  - \nu \alpha -  [x-\bar x(\alpha)]^2 / 2 A^2(\alpha)  }.
\end{equation}
Given a value $x$,
a steepest descent analysis of this integral yields
\begin{equation} \label{e:alpha_p-case-iii}
{\partial \over \partial \alpha} \left. \left(
   \nu f(\alpha) - {(x- \bar x(\alpha))^2 \over 2A^2(\alpha)}\right)
   \right|_{\alpha^*} = \nu,
\end{equation}
as condition for the value $\alpha^*$ maximizing the integrand.
As we vary $x$, the changing value $\alpha^*$ defines a function
$\alpha_P(x)$ which enables us to write the Plischke-R\'acz probability as
\begin{equation} \label{e:pxm-result-iii}
 P(x,M) =
        C_{\rm PR}(x)M^{-\phi(x)}
\end{equation}
where
\begin{equation} \label{e:phi-case-iii}
  \phi(x) = - \nu f(\alpha_P(x)) +  \nu \alpha_P(x) +
   { (x-\bar x(\alpha_P(x)))^2 \over 2A^2(\alpha_P(x)) },
\end{equation}
and $C_{\rm PR}(x)$ is an amplitude.
By comparison of relation (\ref{e:pxm-result-iii}) to
Eq. (\ref{e:pxm-multiscaling}) we see that case (C) like case (B)
results in multiscaling, but with a much more complex multiscaling
``exponent'' $\phi(x)$.

Although the width $\xi(\alpha,M)$ of $N(\alpha,x,M)$ in case (C) still
approaches zero for large $M$, we note that, in contrast to case (B),
$\alpha^*$ is no longer ``characteristic'' for the shell $x$ --- in the
sense that Eq. (\ref{e:alpha-case-ii}) no longer holds. Thus, we refer
to case (C) as a case of ``weak localization.''

To analyze $\rho_A(x,M)$, we first write the condition for $\alpha^*.$ In
analogy to (\ref{e:alpha_p-case-iii}),
\begin{equation} \label{e:alpha_rho-case-iii}
{\partial \over \partial \alpha} \left. \left(
   \nu f(\alpha) - {(x- \bar x(\alpha))^2 \over 2A^2(\alpha)}\right)
   \right|_{\alpha^*} = 0.
\end{equation}
The $\alpha^*$ values satisfying Eq. (\ref{e:alpha_rho-case-iii}) define
a function $\alpha_\rho(x)$, when $x$ is varied.
Unlike the strong localization case (B), $\alpha_\rho(x)$ differs
from $\alpha_P(x)$. We can use the function $\alpha_\rho(x)$ and the
method developed in Eq. (\ref{e:rho-result-i-msform}) to
express $\rho_A(x,M)$ as
\begin{equation} \label{e:rho-result-iii}
    \rho_A(x,M) = C_\rho (x) r ^{D(x)},
\end{equation}
where  $C_\rho (x)$ is an amplitude and the multiscaling exponent
\begin{equation}
   D(x) = f(\alpha_\rho(x)) - {[x-\bar x(\alpha_\rho(x))]^2\over 2\nu
     A^2(\alpha_\rho(x))}.
\end{equation}

The numerical data presented in Sec. \ref{s:simulation} are not
sufficient to distinguish between cases (A) through (C) for all values
of $\alpha$. However, for the particular value $\alpha=1.36$, for which the
statistics is good, we find the data consistent with case (C).
In the following section we will provide further support for the
multiscaling idea.

%
In the preceding discussion we have seen that the existence of two
different length scales, namely the average location $\bar x(\alpha)$
and the width $\xi(\alpha,M$) in cases (B) and (C) leads to multiscaling for
both $P(x,M)$ and $\rho_A(x,M)$. In contrast, we do not find
multiscaling in case (A), where only one length scale is present ---
both $\bar x(\alpha)$ and $\xi(\alpha,M)$ are $M$ independent.

\section{Multifractality}
\label{s:multifractal}

Another interesting quantity --- possibly accessible to experimental
measurements --- is the scaling behavior of ``moments'' of
$N(\alpha,x,M)$. In this section, we will define a multifractal
analysis for the $p_i$ contained in an annulus at distance $x$ in
analogy to the multifractal formalism presented, e.g., in Refs.
\cite{Halsey86a,Stanley88}.

Usually, a multifractal analysis is performed on a set of numbers that
are {\it normalized.} However, the sum of the growth probabilities $p_i$
restricted to an annulus is less than one, since the $p_i$ are
normalized with respect to the entire cluster.  Here, we will
base our analysis on the {\it unnormalized\/} set of $p_i$ at a specific
distance $x$ from the cluster seed and postpone to App. \ref{a:2}
a discussion of what happens if we use normalized probabilities
instead.

Our first step is to define a ``partition function''
\begin{equation} \label{e:zqxm}
 Z(q,x,M) \equiv \int d\alpha N(\alpha,x,M) M^{-q\nu \alpha},
\end{equation}
where $q$ is an arbitrary real number \cite{Jensen91,Schwarzer92}. The function
$Z(q,x,M)$ can also be considered the $q^{\rm th}$ ``moment'' of the
distribution $N(\alpha,x,M)$.  Second, we define the scaling indices
$\tau(q,x)$ as a function of $q$ for different $x$ by
\begin{equation} \label{e:tau}
        \tau(q,x) \equiv \lim_{M\to\infty} {\ln Z(q,x,M) \over \ln M}.
\end{equation}
If $\tau(q,x)$ is a linear function of $q$, then conventional ``gap
scaling'' is obtained, while otherwise we call the measure underlying
the ``moments'' $Z(q,x,M)$ multifractal. It is then convenient to
introduce the Legendre
transform $f_L(\alpha,x)$ of $\tau(q,x)$
\begin{equation} \label{e:legendre}
 f_L (\alpha_L,x) \equiv q\alpha_L - \tau(q,x), \quad {\rm where\ }
    \alpha_L \equiv {\partial\over \partial q}\tau(q,x).
\end{equation}

The quantity $f_L(\alpha_L)$ can be interpreted as the fractal
dimension of the set of points characterized by $\alpha_L$
\cite{Halsey86a} in the annulus $x$.
We analyze the
scaling behavior of the moments by performing a steepest descent
analysis of (\ref{e:zqxm}). Substituting $N(\alpha,x,M)$ in its
analytical form (\ref{e:naxm}) into Eq. (\ref{e:zqxm}) yields
\begin{equation} \label{e:zqxm-analytical}
  Z(q,x,M) =
     \int d\alpha {M^{\nu f(\alpha)-\nu q\alpha} \over
                       \sqrt {2 \pi \xi^2(\alpha,M)} }
     \exp\left[- {(x-\bar x(\alpha))^2 \over 2 \xi^2(\alpha,M)}\right].
\end{equation}
As in Sec. \ref{s:cases} we will now discuss the consequences of
different asymptotic behavior of the width of the growth zone of
the cluster.

\savebox{\boxp}{$\xi(\alpha,M) = A(\alpha)$}
\subsection{Constant Width: \usebox{\boxp}}
\label{s:case-i-2}

For the case of constant width (cf. Sec. \ref{s:case-i}), we find that
the dominant contribution to the integral (\ref{e:zqxm-analytical})
arises at a value $\alpha^*=\alpha^*(1)$ given by
\begin{equation} \label{e:fax-alpha-i}
{d \over d\alpha} f(\alpha)|_{\alpha^*} = q.
\end{equation}
 From Eq. (\ref{e:fax-alpha-i}), we can anticipate the
result,
\begin{equation} \label{e:f_l-case-i}
    f_L(\alpha_L,x) = f(\alpha_L),
\end{equation}
which we obtain after consideration of Eqs. (\ref{e:tau}) and
(\ref{e:legendre}). Eq. (\ref{e:f_l-case-i}) states that {\it for case
(A) the full, unaltered multifractal spectrum is found in all annuli
$x$.}  Thus, given the validity of Eq. (\ref{e:naxm}), a constant width
of the growth zone can only be maintained, if sites with both low
and high growth probabilities are distributed evenly in the cluster
\cite{Mandelbrot90d}.

However, due to the the screening of the
interior regions of the cluster, the growth probabilities for small $x$
are significantly smaller than at the exposed sites on the exterior of
the cluster. These smaller probabilities result in a shift of the
distributions $N(\alpha,x,M)$ to larger values of $\alpha$ for small
$x$. Thus, the multifractal spectrum $f_L(\alpha_L)$ differs for
different $x$ (cf. \cite{Schwarzer92} for the 3D case),
in contrast to the result (\ref{e:f_l-case-i}) above.

\savebox{\boxp}{$\xi(\alpha,M) = A(\alpha) M^{-y}$}
\subsection{Strong Localization: \usebox{\boxp}}
\label{s:case-ii-2}

For the strong localization case (Sec. \ref{s:case-ii}), the Gaussian
term in (\ref{e:zqxm-analytical})  tends to a $\delta$-function
localized at $x$ given by Eq. (\ref{e:alpha-case-ii}) whose inversion
gives $\alpha = \alpha(x)$. Now, we use for $Z(q,x,M)$ in Eq.
(\ref{e:tau}) the value of the integrand in (\ref{e:zqxm}) at
$\alpha=\alpha(x)$. The resulting scaling indices $\tau(q,x)$,
\begin{equation} \label{e:tau-case-ii}
  \tau(q,x) = q\alpha(x) - f(\alpha(x)),
\end{equation}
are linearly dependent on $q$ for fixed $x$. Thus, {\it in case (B) gap scaling
of the
moments $Z(q,x,M)$ results.} The $f_{\rm L}(\alpha_L)$ corresponding to
(\ref{e:tau-case-ii}) is the point $[\alpha_{\rm L} = \alpha(x);~
f_{\rm L}(\alpha) = f(\alpha(x))~]$.

The absence of multifractality in a given annulus $x$ in case (B) is not
surprising, since the strong localization of $\alpha$-sites implies a
very narrow spread of $\alpha$ values within a specific shell $x$.
Narrow distributions, however, typically display gap scaling of their
moments.

\savebox{\boxp}{$\xi(\alpha,M) = A(\alpha) / \sqrt{\ln M}$}
\subsection{Weak Localization: \usebox{\boxp}}
\label{s:case-iii-2}

Again, the weak localization case is quite different. The value
$\alpha^*$ maximizing the integrand of (\ref{e:zqxm}) is given by
\begin{equation}
  {\partial \over \partial\alpha} \left. \left( \nu f(\alpha) - {(x- \bar
  x(\alpha))^2 \over 2A^2(\alpha)}\right) \right|_{\alpha^*} = \nu q.
\end{equation}
In contrast to Eq. (\ref{e:fax-alpha-i}), where the solution depends on
$q$ only, here $\alpha^*$ is a function of both $q$ and $x$,
and we write $\alpha^* = \alpha_q(x)$. From Eq. (\ref{e:tau}), we obtain
\begin{equation} \label{e:tau-case-iii}
  \tau(q,x) = \nu f(\alpha_q(x)) -\nu \alpha_q(x) q -
        {(x-\bar x(\alpha_q(x)))^2\over 2 A^2(\alpha_q(x))}.
\end{equation}
In (\ref{e:tau-case-iii}) $\tau(q,x)$ is a nonlinear function of both
$q$ and $x$, so that its Legendre transform (\ref{e:legendre}) depends
on $x$. Thus, {\it in case (C), the multifractal spectrum
$f_L(\alpha_L,x)$ is a function of the location of the annulus $x$.} In
contrast to case (B), the width $\xi(\alpha,M)$ approaches zero so slow
that the multifractality of case (A) is not destroyed as in case (B),
but altered in its character. In fact, the $x$ dependence of the
multifractal spectrum $f_L(\alpha,x)$ is the hallmark of multiscaling as
encountered for $P(x,M)$ and $\rho_A(x,M)$ in case (C), Sec.
\ref{s:cases}.

In Fig. \ref{f:tau} we display $\tau(q,x)$ for different values of $x$
as a function of $q$ \cite{phase-transition}.

For large $|q|$ the function $\tau(q,x)$ tends to straight lines with
different slopes. The definitions (\ref{e:zqxm},\ref{e:tau}) show that
for $q<0$ the slope is determined by the mass dependence of the smallest
growth probability $p_{\rm min} (x)$ within the annulus $x$ and for
$q>0$ by the mass dependence of the largest growth probability $p_{\rm
max}(x)$, respectively. Especially in the region $0<q<1$, we observe a
pronounced curvature of $\tau(q,x)$.  Since the strong localization case
predicts a linear behavior of $\tau(q,x)$ over the entire range of $q$
values, we conclude that our findings disfavor strong localization.

Moreover, we see that $\tau(q,x)$ displays variation with the parameter
$x$.  If $x$ becomes smaller, both $p_{\rm min}(x)$ and $p_{\rm max}(x)$
as functions of the cluster mass decay faster, because the interior
frozen regions of the cluster are screened stronger.  Thus,
the two asymptotic slopes (for $q\to\pm\infty$) of $\tau(q,x)$ increase.
For example, for any given $q$, $\tau(q,x=1.9)$ has everywhere a slope
less than $\tau(q,x=0.7)$. As a consequence, also the Legendre
transforms of $\tau(q,x)$ are $x$ dependent. This finding is consistent
with the weak localization case (C). However, due to the comparatively
small clusters that we have analyzed, we cannot exclude cases (A) or (B).

In Fig. \ref{f:tau} we see that $\tau(q,x)$ only starts to change
appreciably for quite large $x$-values around $x=1.5$. A similar
phenomenon is observed in the numerical multiscaling analysis of
the annular density of the cluster. There, the function $D(x)$ is
approximately constant $=1/\nu$ up to $x$ values of similar magnitude
before $D(x)$ drops to zero over comparatively small range.

For 3D off-lattice DLA, the multifractal properties of the growth
probabilities $\{p_i\}$ in an annulus were calculated in
\cite{Schwarzer92}. Although no statement about $N(\alpha,x,M)$ for 3D
DLA was made, the results display qualitatively the behavior predicted
above in the cases (A) and (C) for 2D DLA.

\section{conclusion}

We have discussed different possibilities for the analytical form of the
joint distribution function $N(\alpha,x,M).$ For cases (B) and (C),
where two different length scales enter into $N(\alpha,x,M)$, we find
multiscaling behavior of the Plischke-R\'acz probability $P(x,M)$ and
the density profile $\rho_A(x,M)$ of DLA clusters. Moreover, we find
that the scaling behavior of the moments of the growth probability
distribution constrained to an annulus $x$ is different in all cases.
For (A) and (C), we encounter multifractality, which in case (C) bears
an additional feature typical for multiscaling: the $x$ dependence of a
scaling ``exponent.'' Our data and previous work is consistent with case
(C), although further numerical work to clarify the rich
scaling properties of $N(\alpha,x,M)$ is clearly desirable.

Our results for all the different cases discussed above are summarized
in Table \ref{t:1}.

\acknowledgements

We wish to thank P. Ossadnik for helpful interactions and comments on
the manuscript, and NSF for financial support.  One of the authors
(J.L.) thanks the Center for Theoretical Physics at Seoul National
University for financial support and hospitality.

\appendix

\section{Spatial distribution of dead and alive perimeter sites}
\label{a:1}

In this appendix, we discuss the properties of the distribution of dead
perimeter sites \cite{Amitrano89}. These sites with growth probability
exactly equal to zero result because specific configurations of cluster
sites enclose perimeter sites in such a way that they can no longer be
reached from the exterior of the DLA cluster.  If dead sites are
predominantly formed due to specific {\it local\/} configurations, then
the fraction of perimeter sites that are dead will not change as the
cluster grows \cite{Amitrano89}.
Moreover, the spatial distribution of dead sites is then
proportional to the distribution of alive sites, i.e.,
\begin{equation}
        \rho_A^{\rm (d)} (x,M) \sim \rho_A^{\rm (a)}(x,M) \sim \rho_A(x,M).
\end{equation}
On the other hand, if the dead sites were found mainly near the region
of small $p_i$, then $\rho_A^{\rm (d)}(x,M)$ would be shifted towards
the center of the cluster.  In Fig. \ref{f:distributions}, we notice
that the shape of both distributions looks almost identical. The
similarity in form is an evidence favorable to the above stated ``local
configuration argument,'' and shows that the dead sites are uniformly
distributed over the cluster.

The simplest local configuration producing a dead site is the ``${\tt
L}$'' configuration shown in Fig. \ref{f:configurations}a.  In general,
also more complicated configurations will contribute (Fig.
\ref{f:configurations}b--e).  A coarse graining over the scale of lowest
order configurations of this kind is necessary to see that the densities
of dead and alive perimeter sites are proportional.  Reference
\cite{Amitrano89} finds a value ($0.365 \pm 0.01$) for the ratio of the
number of dead perimeter sites to the total number of perimeter sites
for 2D square-lattice DLA.

\section{Consequences of normalization for the multifractal analysis}
\label{a:2}

Here, we will briefly explore the consequences of normalizing the
growth probabilities $p_i$ within an annulus $x$ prior to performing the
multifractal analysis suggested in Sec. \ref{s:multifractal}.
We determine an $x$ and $M$ dependent normalization
factor ${\cal N}(x,M)$ to multiply each $p_i$, such that the sum over
the growth probabilities for fixed $x$ equals $1$.
The normalization procedure alters the value of $\alpha$
associated with each $p_i$ to
\begin{equation} \label{a:a2a}
        \tilde \alpha_i = \log p_i/\log R - \log {\cal N}(x,M) /\log R.
\end{equation}
Now, the distribution of the $\tilde\alpha$ is
$\tilde N(\tilde\alpha,x,M)$, which is related to $N(\alpha,x,M)$ by
\begin{equation}
        \tilde N(\tilde\alpha,x,M) = N(\alpha - \log{\cal N}(x,M)/\log R,x,M).
\end{equation}

In analogy to Eq. (\ref{e:zqxm}) we denote the  $q^{\rm th}$ ``moment'' of
$N(\tilde\alpha,x,M)$ by $\tilde Z(q,x,M)$, i.e.,
\begin{eqnarray}
  \tilde Z(q,x,M) & \equiv & \int d\tilde\alpha \tilde N(\tilde\alpha,x,M)
                        M^{-q\nu\tilde\alpha} \\
                  & = & {1\over \left[{\cal N}(x,M)\right]^q} \int d\alpha
                        \tilde N(\alpha,x,M) M^{-q\nu\alpha} \\
                  & = & {1\over \left[{\cal N}(x,M)\right]^q} Z(q,x,M).
\end{eqnarray}
 From the normalization we know that the first moment of
$\tilde N(\tilde\alpha,x,M)$ is equal to $1$. It follows that
${\cal N}(x,M) = Z(1,x,M)$.

We continue along the lines in Sec. \ref{s:multifractal} and define the
equivalent $\tilde\tau(q,x)$ to $\tau(q,x)$,
\begin{eqnarray}
        \tilde\tau(q,x) & \equiv & - \lim_{R\to\infty}
                \frac{\partial \log \tilde Z(q,x,M)}{\partial \log R} \\
         & =  & \tau(q,x) + q \lim_{R\to\infty}
                \frac{\partial \log{\cal N}(x,M)}{\partial \log R} .
\end{eqnarray}
If the normalization constant ${\cal N}(x,M)$ displays power-law scaling
with the cluster size $R$, then $\tilde\tau(q,x)$ and $\tau(q,x)$ differ
only by a linear function. How does this difference affect the Legendre
transform of $\tilde\tau(q,x)$ when we compare it to the transform of
$\tau(q,x)$ which is defined in Eq. (\ref{e:legendre})?
First, we calculate the slope $\tilde \alpha_L$ of $\tau(q,x)$,
\begin{equation}
   \tilde\alpha_L = \frac{\partial\tilde\tau(q,x)}{\partial q}=
                        \alpha_L + \lim_{R\to\infty}
                        \frac{\partial\log{\cal N}(x,M)}{\partial \log R}.
\end{equation}
Since ${\cal N}(x,M)$ is always less than one, we see
that $\tilde\alpha_L$ is merely $\alpha_L$ shifted by a constant to
smaller values.
Second, Legendre transforming $\tilde\tau(q,x)$ yields
\begin{eqnarray}
   \tilde f_L(\tilde\alpha_L) & = & q\tilde\alpha_L - \tilde\tau(q,x) \\
        & = & q\alpha + q \lim_{R\to\infty}\frac{\partial\log{\cal
                N}(x,M)}{\partial \log R}                         \\
        & & \qquad  - \tau(q,x) - q \lim_{R\to\infty}
                \frac{\partial\log{\cal N}(x,M)}{\partial \log R} \nonumber\\
        & = & q\alpha - \tau(q,x) = f_L(\alpha_L).
\end{eqnarray}
Thus, we retain the functional form of the Legendre transform
$f_L(\alpha_L)$ of $\tau(q,x)$ and the only difference to $\tilde
f(\tilde\alpha_L)$ is that the latter is shifted towards smaller values
of $\alpha$.

For example, if we use the formalism presented in this appendix to
calculate the $\tilde f_L(\tilde\alpha_L)$ for the strong localization
case (cf. Sec. \ref{s:case-ii-2}), we find that
\begin{equation} \label{e:a2-strong-localization}
        \tilde f_L(\tilde\alpha_L) = f_L,\quad {\rm and}\quad
        \tilde\alpha = f_L,
\end{equation}
where $f_L$ denotes the constant value of $f_L(\alpha(x))$ for a given
$x$.  In order to interpret the result (\ref{e:a2-strong-localization}),
consider the normalization condition $\sum \tilde p_i(x) =1$ within an
annulus. In the strong localization case, the annulus $x$ is
characterized by only one $\alpha$. Thus, the the product
$M^{-\nu\tilde\alpha}M^{\nu f(\tilde\alpha)}$ must be constant, leading
to $\tilde\alpha = f_L$.

\widetext          

\begin{table}
\caption{Summary of multifractal and multiscaling features arising
in cases (A), (B) and (C).}

\begin{tabular}{lcccc}
 & $\rho_A(x,M)$ & $P(x,M)$ & \multicolumn{2}{c}{$N(\alpha,x,M)$} \\
 & multiscaling  & multiscaling & multiscaling & multifractal \\
\tableline
(A)~$\xi=A(\alpha)$ & no & no & no & yes \\
(B)~$\xi=A(\alpha)/M^y$ & yes & yes & yes & no \\
(C)~$\xi=A(\alpha)/\sqrt{\ln M}$ & yes & yes & yes & yes \\
\end{tabular}
\label{t:1}
\end{table}

\narrowtext        

\begin{figure}
\caption{
\label{f:configurations}
(a)~ The lowest order configuration which contains a
dead site ($\times$). ~(b-e)~ Examples of higher
order configurations containing dead sites.
}
\end{figure}

\begin{figure}
\caption{
\label{f:dead-alive}
(a) Dead and (b) alive growth sites of a DLA cluster with $M= 20\ 000$,
indicating the similar spatial distribution of both types of perimeter
sites. Dead sites are perimeter sites with $p_i = 0$. Ref.
\protect\cite{Amitrano89} finds that the number of dead perimeter sites in DLA
is proportional to the number of all perimeter sites. In our case,
dead sites constitute a fraction of $\approx 42\%$ of the perimeter sites.
}
\end{figure}

\begin{figure}
\caption{
\label{f:distributions}
Comparison of the two distributions of dead sites, $\rho^{\rm
(d)}_A(x,M)$ (broken line), and of alive sites,
$\rho_A^{(a)}(x,M)$ (solid line). To demonstrate that both are
distributed in a similar fashion, we have scaled $\rho_A^{(d)}(x,M)$ by
the ratio of the number of alive to dead perimeter sites $( =1.36).$
}
\end{figure}


\begin{figure}
\caption{
\label{f:naxm}
$N(\alpha, x, M)$ averaged over $18$ clusters of mass $M = 20\ 000$.
Different symbols denote different values of $\alpha$: $1.1 (\bigcirc),
1.5 (\Box), 1.9 (\bigtriangleup), 2.3 (\bigtriangledown)$ and $2.8
(\bullet)$. To test whether the $x$ dependence of $N(\alpha,x,M)$ can be
represented by a Gaussian, we plot $\pm (x-\langle x\rangle)^2$ on the
abscissa, where the $+$ sign applies if $x \ge \langle x\rangle$ and the
$-$ sign otherwise.  The ordinate scale is logarithmic. Thus, Gaussian
behavior manifests itself in two straight lines emanating from $x- \langle
x\rangle=0$ with slopes of opposite sign but equal magnitude. The two
solid lines in the plot illustrate this behavior and are
intended as guides to the eye for the case $\alpha =1.1$.
}
\end{figure}

\begin{figure}
\caption{
\label{f:alpha-sites}
Location of $\alpha$ sites from $18$ off-lattice DLA clusters of $M=20\ 000$.
In (a)~ we have $1.5 < \alpha < 1.9$, in (b)~ $2.8 < \alpha < 3.0$ and in (c)~
$\alpha > 6$.
}
\end{figure}

\begin{figure}
\caption{
\label{f:width-mean}
(a)~ The width $\xi (\alpha,M)$ and (b)~ the mean position $\langle
x(\alpha,M)\rangle$ vs $M$ for several values of $\alpha$. Different
symbols denote different $\alpha$ values, 0.68 ($\bigcirc$), 1.36
($\Box$), 2.04 ($\bigtriangleup$), 2.72 ($\bigtriangledown$), 3.40
($\bullet$), 4.08 (\protect\rule{5pt}{5pt}), 4.76 (filled upward
triangle) and 5.95 (filled downward triangle).  The data are averaged
over $18$ off-lattice DLA clusters.  In (c)~ $\langle
x(\alpha,M)\rangle$ is plotted as a function of $\alpha$. Here,
different line styles correspond to different cluster masses (see
legend).
}
\end{figure}

\begin{figure}
\caption{
\label{f:tau}
Dependence on $q$ of $\tau(q,x)$ for different values of $x$. Here,
$\tau(q,x)$ was determined by fitting a straight line through the $3$
data points corresponding to $M=5~000,~10~000,~20~000$ in a plot of
$\log Z(q,x,M)$ vs $\log M$ averaged over $18$ DLA clusters.  Different
line styles denote the different $x$ values (legend). For comparison, we
plot the $\tau(q)$ resulting from an analysis of the growth
probabilities of the entire cluster ($\bigcirc$).
Since 2D DLA displays a phase transition \protect\cite{Lee88,Blumenfeld89},
$\tau(q,x)$ for negative $q$ can only be considered an ``effective''
exponent which will display larger and larger slopes as $M\to\infty$.
}
\end{figure}


\begin{references}

\bibitem{Witten81} T. A. Witten and L. Sander, Phys. Rev. Lett. {\bf 47},
1400 (1981).

\bibitem{Turkevich85} L. A. Turkevich and H. Scher, H., Phys.  Rev.  Lett.
{\bf 55}, 1026 (1985).

\bibitem{Ball85} R. C. Ball, R. M. Brady, G. Rossi  and B. R. Thompson,
Phys. Rev. Lett. {\bf 55}, 1406 (1985).

\bibitem{Coniglio85} A. Coniglio, in {\it On Growth and Form: Fractal and
Non-Fractal Patterns in Physics\/}, H. E.  Stanley and N.
Ostrowsky (eds) (Nijhoff, Dordrecht, 1985), p. 101.

\bibitem{Parisi85} G. Parisi and Y. C. Zhang, J. Stat.  Phys. {\bf 41}, 1
(1985).

\bibitem{Hayakawa87} Y. Hayakawa, S. Sato and M. Matsushita, M., Phys.  Rev. A
{\bf 36}, 1963 (1987).

\bibitem{Halsey87} T. C. Halsey, Phys. Rev. Lett. {\bf 59}, 2067
(1987).

\bibitem{Pietronero88} L. Pietronero, A. Erzan and C. J. G. Evertsz, Phys.
Rev.  Lett.  {\bf 61}, 861 (1988).

\bibitem{Meakin88} P. Meakin, in {\it Phase Transitions and Critical
Phenomena} (eds. C. Domb and J. L. Lebowitz), Vol. 12 (Academic,
Orlando, 1988).

\bibitem{b:Feder88} J. Feder, {\it Fractals} (Pergamon, NY, 1988).

\bibitem{b:Stanley88} H. E. Stanley and N. Ostrowsky (eds), {\it Random
Fluctuations and Pattern Growth: Experiments and Models\/} (Kluwer
Academic Publishers, Dordrecht, 1988).

\bibitem{b:Pietronero90}
L. Pietronero (ed), {\it Fractals: Physical Origin and
Properties\/} (Plenum Publishing Co., London, 1990) [Proc. 1988 Erice
Workshop on Fractals].

\bibitem{b:Bunde91}
A. Bunde and S. Havlin (eds), {\it Fractals and Disordered Systems\/}
(Springer, Berlin, 1991).

\bibitem{b:Vicsek92} T. Vicsek, {\it Fractal Growth Phenomena\/}
(World, Singapore, 1989).

\bibitem{Meakin85} P. Meakin, H. E. Stanley, A. Coniglio and T. A. Witten,
Phys. Rev. A {\bf 32}, 2364 (1985).

\bibitem{Meakin86} P. Meakin, A. Coniglio, H. E. Stanley and T. A. Witten,
Phys. Rev. A {\bf 34}, 3325 (1986).

\bibitem{Halsey86}
T. C. Halsey, P. Meakin and I. Procaccia, Phys. Rev. Lett.
{\bf 56}, 854 (1986).

\bibitem{Amitrano86} C. Amitrano, A. Coniglio and F. di Liberto,  Phys. Rev.
Lett. {\bf 57}, 1016 (1986).

\bibitem{Amitrano89} C. Amitrano, P. Meakin and H. E. Stanley, Phys. Rev. A
{\bf 40}, 1713 (1989).

\bibitem{Halsey86a}
T. C. Halsey, M. H. Jensen, L. P. Kadanoff, I. Procaccia and B. I. Shraiman,
Phys. Rev. A {\bf 33}, 1141 (1986).

\bibitem{Stanley88}
 H. E. Stanley and P. Meakin, Nature {\bf 335}, 405(1988).

\bibitem{Mandelbrot90} B.B. Mandelbrot, C.J.G. Evertsz and Y. Hayakawa, Phys.
Rev. A {\bf 42},  (1990) 4528.

\bibitem{Mandelbrot91} B.B. Mandelbrot and C.J.G. Evertsz, Physica A
{\bf 177}, 386 (1991).

\bibitem{Schwarzer91}
S. Schwarzer, J. Lee, S. Havlin, H. E. Stanley and P. Meakin,
Phys. Rev. A {\bf 43}, 1134 (1991).

\bibitem{Plischke84} M. Plischke and Z. R\'acz, Phys. Rev. Lett.
{\bf 53}, 415 (1984).

\bibitem{Meakin85a} P. Meakin and L. M. Sander, Phys. Rev. Lett. {\bf 54},
2053 (1985).

\bibitem{Coniglio90} A. Coniglio and M. Zannetti, Physica A {\bf 163}, 325
(1990).

\bibitem{Amitrano91} C. Amitrano, A. Coniglio, P. Meakin and M. Zanetti,
Phys. Rev. {\bf 44}, 4974 (1991).

\bibitem{PO-private} P. Ossadnik, Physica A, in press.

\bibitem{Jensen91} In the context of turbulence a connection between
a sort of multiscaling and multifractality has been discussed by
M. H. Jensen, G. Paladin and A. Vulpiani, Phys. Rev. Lett. {\bf 67}, 208
(1991).

\bibitem{Meakin84} P. Meakin, I. Majid, S. Havlin and H. E. Stanley, J.
Phys. A {\bf 17}, L975(1984).

\bibitem{generalization} a more general form for $N(\alpha,x,M)$
would include a power-law of $x$ multiplying the pure Gaussian behavior
\protect\cite{Ossadnik92a}. However, such an $x$-dependent term would
not invalidate the discussion in the subsequent sections and so we
adhere to the Gaussian form for the purposes of this paper.

\bibitem{Makarov85} N. G. Makarov, Proc. London Math. Soc. {\bf 51}, 369
(1985).

\bibitem{Ossadnik92a} P. Ossadnik and J. Lee, J. Phys. A, submitted.

\bibitem{phase-transition}
Since the small growth probabilites in 2D DLA vanish faster than any
power of $M$, DLA displays a phase transition in its
multifractal spectrum.
This phase transition manifests itself in
increasing slopes of $\tau(q,x)$ for negative $q$, when $\tau(q,x)$ is
estimated from clusters with increasingly larger masses $M$.
We expect it to occur for all $x$ for which $\tau(q,x)$ is defined,
since Ref. \protect\cite{Mandelbrot90d} has demonstrated that the small
growth probabilties are spatially distributed over the entire cluster.

\bibitem{Lee89} J. Lee, P. Alstr\o m and H. E. Stanley, Phys. Rev.  A {\bf 39},
6545 (1989).

\bibitem{Schwarzer92}
For 3D off-lattice DLA, the multifractal spectrum in a shell around the
cluster seed was investigated in S. Schwarzer, S. Havlin
and H. E. Stanley, Physica A {\bf 191}, 117 (1992).

\bibitem{Ossadnik91} P. Ossadnik, Physica A {\bf 176}, 454 (1991).

\bibitem{Mandelbrot90d}
B. B. Mandelbrot and C. J. G. Evertsz, Nature {\bf 348}, 143 (1990).

\bibitem{Lee88} J. Lee and H. E. Stanley, Phys. Rev. Lett.
{\bf 61}, 2945 (1988).

\bibitem{Blumenfeld89}  R. Blumenfeld and A. Aharony, Phys. Rev. Lett.
{\bf 62}, 2977 (1989).

\end{references}
\end{document}